\documentclass[aip,apl,reprint]{revtex4-1}
\usepackage[T1]{fontenc}
\usepackage[latin9]{inputenc}
\usepackage{amstext}
\usepackage{graphicx}
\usepackage{amssymb}
\usepackage{multirow}


\begin{document}

\preprint{}

\title{Multiplexed dispersive readout of superconducting phase qubits}

\author{Yu Chen}
\affiliation{Department of Physics, University of California, Santa Barbara CA 93106, USA}
\author{D. Sank}
\affiliation{Department of Physics, University of California, Santa Barbara CA 93106, USA}
\author{P. O'Malley}
\affiliation{Department of Physics, University of California, Santa Barbara CA 93106, USA}
\author{T. White}
\affiliation{Department of Physics, University of California, Santa Barbara CA 93106, USA}
\author{R. Barends}
\affiliation{Department of Physics, University of California, Santa Barbara CA 93106, USA}
\author{B. Chiaro}
\affiliation{Department of Physics, University of California, Santa Barbara CA 93106, USA}
\author{J. Kelly}
\affiliation{Department of Physics, University of California, Santa Barbara CA 93106, USA}
\author{E. Lucero}
\affiliation{Department of Physics, University of California, Santa Barbara CA 93106, USA}
\author{M. Mariantoni}
\affiliation{Department of Physics, University of California, Santa Barbara CA 93106, USA}
\affiliation{California NanoSystems Institute, University of California, Santa Barbara CA 93106, USA}
\author{A. Megrant}
\affiliation{Department of Materials, University of California, Santa Barbara CA 93106, USA}
\author{C. Neill}
\affiliation{Department of Physics, University of California, Santa Barbara CA 93106, USA}
\author{A. Vainsencher}
\affiliation{Department of Physics, University of California, Santa Barbara CA 93106, USA}
\author{J. Wenner}
\affiliation{Department of Physics, University of California, Santa Barbara CA 93106, USA}
\author{Y. Yin}
\affiliation{Department of Physics, University of California, Santa Barbara CA 93106, USA}
\author{A. N. Cleland}
\affiliation{Department of Physics, University of California, Santa Barbara CA 93106, USA}
\affiliation{California NanoSystems Institute, University of California, Santa Barbara CA 93106, USA}
\author{John M. Martinis}
\affiliation{Department of Physics, University of California, Santa Barbara CA 93106, USA}
\affiliation{California NanoSystems Institute, University of California, Santa Barbara CA 93106, USA}

\begin{abstract}
We introduce a frequency-multiplexed readout scheme for superconducting phase qubits. Using a quantum circuit with four phase qubits, we couple each qubit to a separate lumped-element superconducting readout resonator, with the readout resonators connected in parallel to a single measurement line. The readout resonators and control electronics are designed so that all four qubits can be read out simultaneously using frequency multiplexing on the one measurement line. This technology provides a highly efficient and compact means for reading out multiple qubits, a significant advantage for scaling up to larger numbers of qubits.
\end{abstract}

\maketitle

Quantum computers can execute certain algorithms exponentially faster than their classical counterparts, in particular Shor's factoring algorithm.\cite{nielsen2000} This is achieved by creating complex superposition states of multiple quantum bits (qubits), with the computational advantage over classical methods appearing for large numbers of qubits. Superconducting approaches to quantum computing hold great promise, due to the relatively good performance displayed by superconducting qubits, combined with the ease with which complex superconducting integrated circuits can be fabricated. This is witnessed by the number of recent publications describing beautiful multi-qubit experiments implemented in superconducting architectures.\cite{mariantoni2011, reed2012, fedorov2012, dicarlo2009, lucero2012}

Scaling up to larger numbers of qubits is however a significant challenge, in part because each qubit must be separately controlled and measured. Any simplification or reduction in the resources needed to implement control or measurement would provide a important advantage. In this Letter we describe a multiplexed qubit readout scheme for the superconducting phase qubit that promises highly efficient scaling, a scheme that is also applicable to other qubit systems.

\begin{figure}
\includegraphics{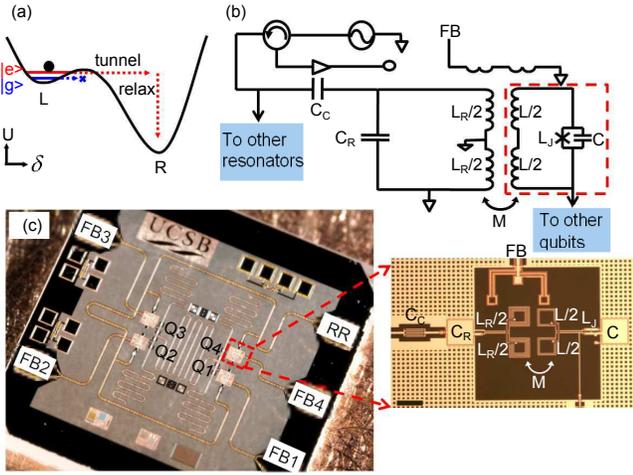}
\caption{(Color online) (a) Schematic representation of qubit projective measurement, where a current pulse allows a qubit in the excited state $|e\rangle$ to tunnel to the right well ($R$), while a qubit in the ground state $|g\rangle$ stays in the left well ($L$). (b) Readout circuit, showing lumped-element $L_R-C_R$ readout resonator inductively coupled to the qubit, with Josephson junction effective inductance $L_J$ and capacitance $C$, with loop inductance $L$. Qubit control is through the differential flux bias line ($FB$). The readout resonator is capacitively coupled through $C_c$ to the readout line, in parallel with the other readout resonators. The readout line is connected through a cryogenic circulator to a low-noise cryogenic amplifier and to a room temperature microwave source. (c) Photomicrograph of four-qubit sample.  $FB1-4$ are control lines for each qubit and $RR$ is the resonator readout line. Inset shows details for one qubit and its readout resonator. Scale bar is 50 $\mu m$ in length; \label{fig.setup}}
\end{figure}

A phase qubit's quantum state is measured by the process illustrated in Fig.\,\ref{fig.setup}(a). A current pulse is applied to the qubit, lowering the barrier between the metastable computational energy well (marked $L$ for left well) and the minimum energy well (marked $R$ for right well). The qubit excited state $|e\rangle$ selectively tunnels from the left into the right well, where its energy relaxes, while the ground state $|g\rangle$ stays in the left well. Following this projective measurement, the outcome is determined by reading out in which well the qubit resides. The readout is typically done using a superconducting quantum interference device (SQUID), which can distinguish between the values of magnetic flux that correspond to the left and right wells, and can thus identify whether the qubit was projected onto the $|g\rangle$ or the $|e\rangle$ state.

This readout scheme has high single-shot fidelity (typically better than 90\%), and protects the qubit from dissipative effects associated with some other readout schemes.\cite{neeley2008} However there are some deleterious effects, including the generation of excess quasiparticles from switching to the voltage state of the SQUID\cite{lenander2011}, and the need for more complex fabrication, as the readout circuits involve fabricating three SQUID Josephson junctions for each qubit. To overcome some of these problems, dispersive microwave readout schemes have been developed for the phase qubit,\cite{steffen2010, wirth2010} in which either the qubit itself or an adjacent SQUID modulates the scattering parameters of a nearby microwave transmission line. These techniques eliminate the generation of quasiparticles, but limit the qubit performance by adding decoherence from the direct connection to the transmission line, or still require SQUID co-fabrication, respectively.

Here, we replace the readout SQUID with a lumped-element readout resonator that is weakly coupled to the qubit and to a nearby readout transmission line, as shown in Fig.\,\ref{fig.setup}(b). This significantly simplifies fabrication, eliminates quasiparticles and other heating effects, and maintains good qubit coherence. The projective measurement is the same as that of the SQUID readout, but the result is read out by measuring the effective inductance of the qubit Josephson junction. Each qubit is coupled to its readout resonator through a mutual inductance $M$, such that it presents an effective parallel inductance $\Delta L$ to the resonator, given by
\begin{equation}\label{eq.qubitind}
    \Delta L = \frac{-M^2}{L+\alpha L_J},
\end{equation}
where $L$ is the qubit loop inductance, $L_{J}$ is the effective Josephson inductance and $\alpha = 1 - \omega^2/\omega_0^{2} \approx 0.8 $ is a detuning factor that depends on the microwave probe frequency $\omega$. For the device here, $\omega \approx 4$ GHz and $\omega_0 = 1/\sqrt{LC} \approx 6.5$ GHz, determined by the qubit parallel inductance $L$ and capacitance $C$. The effective Josephson inductance $L_{J}$ has a different value when the qubit is in the left or in the right well. The readout resonator frequency is $\omega = 2 \pi f = 1/\sqrt{C_R (L_R + \Delta L)}$, and thus depends on the qubit state through $L_J$, so the phase of a probe signal reflected off the resonator will also depend on whether the qubit is in the left or right well.

We demonstrated the multiplexed readout using a quantum circuit comprising four phase qubits and five integrated resonators, shown in Fig.\,\ref{fig.setup}(c). The design of this chip is similar to that used for a recent implementation of Shor's algorithm,\cite{lucero2012}, but here the qubits were read out off a single line using microwave reflectometry, replacing the SQUID readout used Ref 6. . This dramatically simplifies the chip design and significantly reduces the footprint of the quantum circuit. We designed the readout resonators so that they resonated at frequencies of 3-4 GHz (far de-tuned from the qubit $|g\rangle \leftrightarrow |e\rangle$ transition frequency of 6-7 GHz), with loaded resonance linewidths of a few hundred kilohertz. This allows us to use frequency multiplexing, which has been successfully used in the readout of microwave kinetic inductance detectors\cite{mazin2006, yates2009} as well as other types of qubits.\cite{jerger2011,jerger2012} Combined with custom GHz-frequency signal generation and acquisition boards, this approach provides a compact and efficient readout scheme that should be applicable to systems with 10-100 qubits using a single readout line, with sufficient measurement bandwidth for microsecond-scale readout times.

As the readout was performed by measuring the reflection off a single transmission line, some care was taken in the microwave design in order to avoid standing waves from reflections. This included designing the coupling capacitor values and adjusting the lengths of the readout lines between successive resonators; for example, the readout line has an extra bend between qubits $Q2$ and $Q3$.

\begin{figure}
\includegraphics{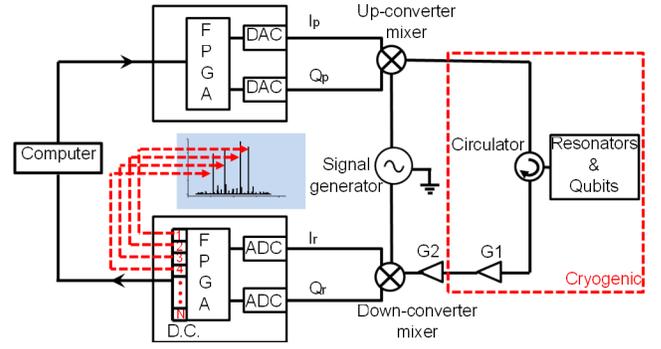}
\caption{(Color online) Setup for frequency-multiplexed readout. Multiplexed readout signals $I_p$ and $Q_p$ from top FGPA-DAC board are up-converted by mixing with a fixed microwave tone, then pass through the circulator into the qubit chip. Reflected signals pass back through the circulator, through the two amplifiers $G1$ and $G2$, and are down-converted into $I_r$ and $Q_r$ using the same microwave tone, and are then processed by the bottom ADC-FPGA board. Data in the shadowed region are the down-converted $I_r$ and $Q_r$ spectra output from the ADC-FPGA board; probe signals from the FPGA-DAC board have the same frequency spectrum. $DC$ indicates the digital demodulation channels, each processed independently and sent to the computer.
\label{fig.measure}}
\end{figure}

We used a standard heterodyne detection method to measure the reflected signal from each readout resonator, as shown in Fig.\,\ref{fig.measure}. Key components include two customized field-programmable gate array (FPGA) boards, one connected to a 14 bit digital-to-analog converter (DAC) for generating arbitrary probe waveforms, with a 1 GS/s digitizing rate (GS/s: gigasample per second), and the other connected to a 8 bit analog-to-digital converter (ADC) for data acquisition and processing, also with a 1 GS/s digitizing rate. Probe waveforms were generated by preparing multi-tone signals in both $I_p$ and $Q_p$ (cosine and sine) probe quadratures for mixer up-conversion, each tone chosen so that after frequency up-conversion in an $IQ$ mixer, it matched the resonance frequency $f_n$ ($n=1-4$ corresponding to the readout resonator for qubits 1-4) of the readout resonators. The reflected signals were amplified and down-converted with a second $IQ$ mixer; the reflected $I_r(t)$ and $Q_r(t)$ signals comprise the same signal tones as the probe waveform, but with an additional phase shift that encodes the measurement signal, i.e. the phase $\phi_n$ of the reflected tone at frequency $f_n$ encodes the state of qubit $n$. The phases $\phi_n$, $n=1-4$, were then evaluated using the digital demodulation channel on the data acquisition FPGA board. This is performed by digital mixing and integrating of the digitized $I_r(t)$ and $Q_r(t)$ signals at the resonator frequency $f_n$, each with a separate demodulation channel.

\begin{figure}
\includegraphics{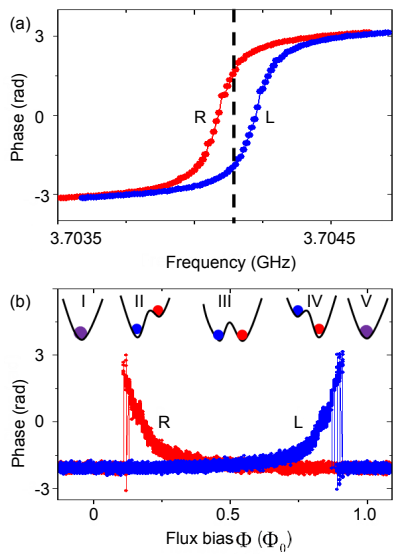}
\caption{(Color online) (a) Phase of signal reflected from readout resonator, as a function of the probe microwave frequency (averaged 900 times), for the qubit in the left ($L$, blue) and right ($R$, red) wells. Dashed line shows probe frequency for maximum visibility. (b) Reflected phase as a function of qubit flux bias, with no averaging. See text for details. \label{fig.phase}}
\end{figure}

The calibration of the readout process was done in two steps. We first optimized the microwave probe frequency to maximize the signal difference between the left and right well states. This was performed by measuring the reflected phase $\phi$ as a function of the probe frequency, with the qubit prepared first in the left and then in the right well. In Fig.\,\ref{fig.phase}(a), we show the result with the qubit flux bias set to 0.15 $\Phi_0$, where the difference in $L_J$ in two well states was relatively large. The probe frequency that maximized the signal difference was typically mid-way between the loaded resonator frequencies for the qubit in the left and right wells, marked by the dashed line in Fig.\,\ref{fig.phase}(a). We typically obtained resonator frequency shifts as large as $\sim150$ kHz for the qubit between the two wells, as shown in Fig.\,\ref{fig.phase}(a), significantly larger than the resonator linewidth.

With the probe frequency set in the first step, the flux bias was then set to optimize the readout. As illustrated in Fig.\,\ref{fig.phase}(b), the optimization was performed by measuring the resonator's reflected phase as a function of qubit bias flux, at the optimal probe frequency, 3.70415 GHz in this case.  The qubit was initialized by setting the flux to its negative ``reset'' value (position I), where the qubit potential has only one minimum. The flux was then increased to an intermediate value $\Phi$, placing the qubit state in the left well, and the reflection phase measured with a 5 $\mu$s microwave probe signal (blue data). The flux was then set to its positive reset value (position V), then brought back to the same flux value $\Phi$, placing the qubit state in the right well, and the reflection phase again measured with a probe signal (red data). Between the symmetry point III ($\Phi = 0.5$) and the regions with just one potential minimum ( $ \Phi \leq 0.1$ or $\Phi \geq 0.9$), the qubit inductance differs between the left and right well states, which gives rise to the difference in phase for the red and blue data measured at the same flux. This difference increases for the flux bias closer to the single-well region, which can give a signal-to-noise ratio as high as 30 at ambient readout microwave power. The optimal flux bias was then set to a value where the readout had a high signal-to-noise ratio (typically $>$ 5), but with a potential barrier sufficient to prevent spurious readout-induced switching between the potential wells. Several iterations were needed to optimize both the probe frequency and flux bias.

Using the optimal probe frequency and flux bias, this readout scheme can distinguish between $|g\rangle$ and $|e\rangle$ with a measurement fidelity of about 90\%, somewhat less than we typically achieve with a SQUID readout. The qubit energy decay time was measured to be $T_1 \approx 600$ ns and the dephasing time $T_{\phi} \approx 150$ ns, for all the qubits. These are typical values for our phase qubits, indicating that replacing the readout SQUID with a resonator did not introduce any significant additional decoherence.

\begin{figure}
\includegraphics{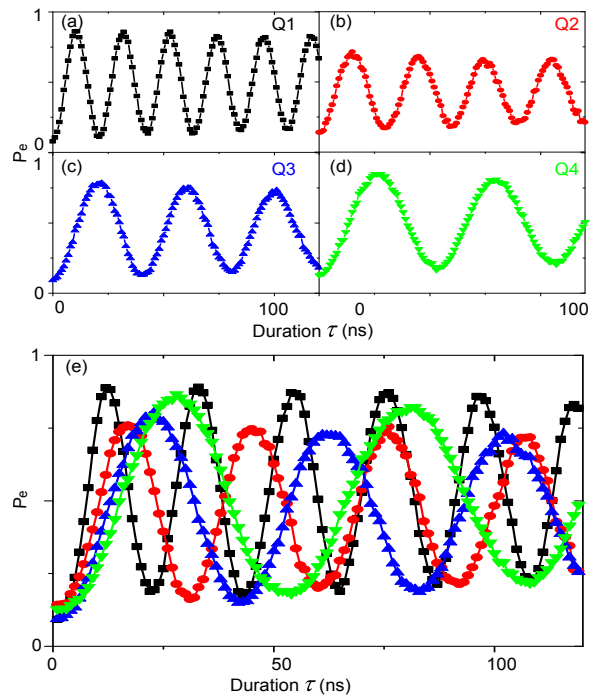}
\caption{(Color online) (a)-(d) Rabi oscillations for qubits $Q1$-$Q4$ respectively, with the qubits driven with 1, 2/3, 1/2 and 2/5 the on-resonance drive amplitude needed to perform a 10 ns Rabi $|g\rangle \rightarrow |e\rangle$ transition. (e) Rabi oscillations measured simultaneously for all the qubits, using the same color coding and drive amplitudes as for panels (a)-(d). \label{fig.rabi}}
\end{figure}

With the bias points chosen for each qubit, we demonstrated the frequency-multiplexed readout by performing a multi-qubit experiment. To minimize crosstalk, we removed the coupling capacitors between qubits used in Ref. 6. In this experiment, we drove Rabi oscillations on each qubit's $|g\rangle \leftrightarrow |e\rangle$ transition and read out the qubit states simultaneously. We first calibrated the pulse amplitude needed for each qubit to perform a $|g\rangle \rightarrow |e\rangle$ Rabi transition in 10 ns. The drive amplitude was then set to 1, 2/3, 1/2 and 2/5 the calibrated Rabi transition amplitude for qubits $Q1$ to $Q4$ respectively, so that the Rabi period was 20 ns, 30 ns, 40 ns and 50 ns for qubits $Q1$ to $Q4$. We then drove each qubit separately using an on-resonance Rabi drive for a duration $\tau$, followed immediately by a projective measurement and qubit state readout. This experiment yielded the measurements shown in Fig.\,\ref{fig.rabi}(a)-(d) for qubits $Q1$-$Q4$ respectively.

With each qubit individually characterized, we then excited and measured all four qubits simultaneously, as shown in Fig.\,\ref{fig.rabi}(e). There is no measurable difference between the individually-measured Rabi oscillations in panels (a)-(d) compared to the multiplexed readout in panel (e).

In summary, we have demonstrated a frequency-multiplexed qubit readout scheme for superconducting phase qubits. Using a single excitation and readout line, with a single amplifier chain, we can measure four qubits simultaneously, without sacrificing measurement bandwidth or qubit coherence. This technology can be scaled up to readout simultaneously tens to hundreds of qubits, greatly aiding the scaling-up of quantum circuits to larger numbers of qubits.

\textbf{Acknowledgements.} This work was supported by IARPA under ARO Award No. W911NF-08-01-0336 and under ARO Award No. W911NF-09-1-0375. M.M. acknowledges support from an Elings Postdoctoral Fellowship. R.B. acknowledges support from the Rubicon program of the Netherlands Organisation for Scientific Research. Devices were made at the UC Santa Barbara Nanofabrication Facility, a part of the NSF funded National Nanotechnology Infrastructure Network.

\end{document}